\documentclass[conference]{IEEEtran}
\IEEEoverridecommandlockouts
\usepackage{cite}
\usepackage{amsmath,amssymb,amsfonts}
\usepackage{algorithmic}
\usepackage{graphicx}
\usepackage{textcomp}
\usepackage{xcolor}
\def\BibTeX{{\rm B\kern-.05em{\sc i\kern-.025em b}\kern-.08em
    T\kern-.1667em\lower.7ex\hbox{E}\kern-.125emX}}
\begin{document}

\title{Emotion Analysis on EEG Signal Using Machine Learning and Neural Network 
}

\author{

\IEEEauthorblockN{S. M. Masrur Ahmed}
\IEEEauthorblockA{\textit{Software Engineer} \\
\textit{bKash Limited}\\
Dhaka, Bangladesh\\
s.m.masrur.ahmed@g.bracu.ac.bd }
\and
\IEEEauthorblockN{Eshaan Tanzim Sabur}
\IEEEauthorblockA{\textit{Department of Computer Science} \\
\textit{BRAC University}\\
Dhaka, Bangladesh\\
eshaan.tanzim.sabur@g.bracu.ac.bd }

}

\maketitle
\begin{abstract}
Emotion has a significant influence on how one thinks and interacts with others. It serves as a link between how a person feels and the actions one takes, or it could be said that it influences one's life decisions on occasion. Since the patterns of emotions and their reflections vary from person to person, their inquiry must be based on approaches that are effective over a wide range of population regions. To extract features and enhance accuracy, emotion recognition using brain waves or EEG signals requires the implementation of efficient signal processing techniques. Various approaches to human-machine interaction technologies have been ongoing for a long time, and in recent years, researchers have had great success in automatically understanding emotion using brain signals. In our research, several emotional states were classified and tested on EEG signals collected from a well-known publicly available dataset, the DEAP Dataset, using SVM (Support Vector Machine), KNN (K-Nearest Neighbor), and an advanced neural network model, RNN (Recurrent Neural Network), trained with LSTM (Long Short Term Memory). The main purpose of this study is to improve ways to improve emotion recognition performance using brain signals. Emotions, on the other hand, can change with time. As a result, the changes in emotion over time are also examined in our research.
\end{abstract}


\section{Introduction}
Emotion is defined as a person's conscious or unconscious behavior that indicates our response to a situation. Emotion is interconnected with a person's personality, mood, thoughts, motivation, and a variety of other aspects. Fear, happiness, wrath, pride, anger, panic, despair, grief, joy, tenseness, surprise, confidence, enthusiasm are the common emotions are all experienced by humans \cite{b1}. The experience can be both positive or negative. In the light of this, physiological indications such as heart rate, blood pressure, respiration signals, and Electroencephalogram (EEG) signals might be useful in properly recognizing emotions.Emotion recognition has always been a major necessity for humanity, not just for usage in fields like computer science, artificial intelligence, and life science, but also for assisting those who require emotional support. For a long time, experts couldn't figure out a reliable way to identify true human emotion. One method was to use words, facial expression, behavior, and image to recognize one's emotions \cite{b2,b3,b4,b5}. Researchers found that subject answers are unreliable for gauging emotion; people are unable to reliably express the strength and impact of their feelings. Furthermore, it is simple to manipulate self-declared emotions, resulting in incorrect findings. As a result, researchers had to shift their focus to approaches that do not rely on subject reactions. The development of Brain-Computer Interface (BCI) and Electroencephalogram (EEG) signals demonstrated more accurate methods for detecting human emotions. It introduced an involuntary approach to get more accurate and reliable results. Involuntary signals are uncontrollable and detect people's true feelings. They have the ability to express genuine emotions. The advancement of a reliable human emotion recognition system using EEG signals could help people regulate their emotions and open up new possibilities in fields like education, entertainment, and security and might aid people suffering from Alexithymia or any other psychiatric disease.
The goal of our paper is to use effective techniques on DEAP dataset to extract features from EEG signals using band waves and apply machine learning algorithms and neural network models to check the efficiency of the used algorithms on valence-arousal, EEG regions and band waves.
% \begin{figure}[htbp]
% \centerline{\includegraphics{BRAIN.jpg}}
% \caption{Lobes of the brain}
% \label{fig}
% \end{figure}

\section{Literature Review}

The EEG research community is expanding its reach into a number of different fields.
In her research, Vanitha V. et al. \cite{b6} aims to connect stress and EEG, and how stress can have both beneficial and bad effects on a person's decision-making process. She also discusses how stress affects one's interpersonal, intrapersonal, and academic performance and argues that stress can cause insomnia, lowered immunity, migraines, and other physical problems.
Jin et al. \cite{b7} while analyzing emotions reported promising results, claiming that combining FFT, PCA, and SVM yielded results that were about 90 percent accurate. As a result, rather than the complexity of the classification algorithm used, the feature extraction stage determines the accuracy of any model. As a result, categorization systems can offer consistent accuracy and recall.
Liu et al. \cite{b8} proposed a fractal-based algorithm to identify and visualize emotions in real time. They found that gamma band could be used to classify emotion. For emotion recognition, the authors analyzed different kinds of EEG features to find the trajectory of changes in emotion. They then proposed a simple method to track the changes in emotion with time. In this paper, the authors built a bimodal deep auto encoder and a single deep auto encoder to produce shared representations of audios and images. They also explored the possibility of recognizing emotion in physiological signals. Two different fusion strategies were used to combine eye movement and EEG data. The authors tested the framework for cross modal learning tasks. The authors introduce a novel approach that combines deep learning and physiological signals.
The DEAP Dataset was also utilized by the following writers to analyze emotion states.
Xing et al. \cite{b9} developed a stacked autoencoder (SAE) to breakdown EEG data and classify them using an LSTM model. - The observed valence accuracy rate was 81.1 percent, while the observed arousal accuracy rate was 74.38 percent.
Chao et al. \cite{b10} investigated a deep learning architecture, reaching an arousal rate of 75.92 percent. and 76.83 percent for valence states.
Mohammadi et al. \cite{b11} classified arousal and valence using Entropy and energy of each frequency band and reached an accuracy of 84.05 percent for arousal and 86.75 percent for valence.
Xian et al. \cite{b12} utilized MCF with statistical, frequency, and nonlinear dynamic characteristics to predict valence and arousal with 83.78 percent and 80.72 percent accuracy, respectively.
Ang et al. \cite{b13} developed a wavelet transform and time-frequency characteristics with ANN classification method. For joyful feeling, the classification rate was 81.8 percent for mean and 72.7 percent for standard deviation y. The performance of frequency domain characteristics for sad emotions was 72.7 percent.
Alhagry et al. \cite{b14} developed a deep learning technique for identifying emotions from raw EEG data that used long-short term memory (LSTM) neural networks to learn features from EEG signals and then classified these characteristics as low/high arousal, valence, and liking. The DEAP data set was used to evaluate the -e technique. -The method's average accuracy was 85.45 percent for arousal and 85.65 percent for valence.
\section{Methodology}
\subsection{Data Materials}
For our research, we have chosen the DEAP \cite{b15} dataset. The DEAP dataset for emotion classification is freely available on the internet. A number of physiological signals found in the DEAP dataset can be utilized to determine emotions. It includes information on four main types of states: valence, arousal, dominance, and liking. Due to the use of various sample rates and different types of tests in data gathering, the DEAP Dataset is an amalgamation of many different data types. EEG data was gathered from 32 participants, comprising 16 males and 16 women, in 32 channels. The EEG signals were collected by playing 40 different music videos, each lasting 60 seconds, and recording the results. Following the viewing of each video, participants were asked to rate it on a scale of one to nine points. According to the total number of video ratings received, which was 1280, the number of videos (40) multiplied by the number of volunteers (40) yielded the result (i.e. 32). Following that, the signals from 512 Hz were downsampled to 128 Hz and denoised utilizing bandpass and lowpass frequency filters, as well as a lowpass frequency filter. 512 Hz EEG signals were acquired from the following 32 sensor positions (using the worldwide 10- 20 positioning system): Fp1, AF3, F3, F7, FC5, FC1, C3, T7, CP5, CP1, P3, P7, PO3, O1, Oz, Pz, Fp2, AF4, Fz, F4, F8, FC6, FC2, Cz, T8, CP2, P4, P8, PO4, and O2. It was also possible to take a frontal face video of each of the 22 participants. Several signals, including EEG, electromyograms, breathing region, plethysmographs, temperature, and so on, were gathered as 40 channel data during each subject's 40 trials, with each channel representing a different signal. EEG data is stored in 32 of the 40 available channels. The rest of the channels record EOG, EMG, ECG, GSR, RSP, TEMP and PLET data.
\subsection{Data Visualization}
We extracted valence and arousal ratings from the dataset. The combination of Valence and Arousal can be converted to emotional states: High Arousal Positive Valence (Excited, Happy), Low Arousal Positive Valence (Calm, Relaxed), High Arousal Negative Valence (Angry, Nervous) and Low Arousal Negative Valence (Sad, Bored). We have analyzed the changes in emotional state along with the number of trials for each group by following Russell’s circumplex model.
Russell's circumplex model helped classify the DEAP dataset. Russell's methodology for visualizing the scale with the real numbers, the DEAP dataset employs self-assessment manikins (SAMs) \cite{b16}. 1–5 and 5–9 were chosen as the scales based on self-evaluation ratings \cite{b17,b18,b19}. The label was changed to “positive” if the rating was greater than or equal to 5, and to “negative” if it was less than 5.
We utilized a different way to determine "positive" and "negative" values. The difference in valence and arousal was rated on a scale of 1 to 9 by the participants of DEAP. We believe that categorizing the dataset using a mean value is not a good approach because there may be no participants who rate between 1-2 and 4-6. As a result, using a mean value to derive the separation could lead to bias. On the other hand, all users may have given ratings ranging from 5 to 9. To avoid biased analysis, we wanted to utilize the value from the mid range to separate the positive and negative values. As a result, to distinguish between "positive" and "negative" numbers, we used median values. We looked for a positive or negative valence as well as a positive or negative arousal level in each experiment. Numbers greater than the median are considered "positive", while values less than the median are considered "negative".
\newline
Four labels for our research have been created: high arousal low valence (HALV), low arousal high valence (LAHV), high arousal high valence (HAHV), and low arousal low valence (LALV).

\subsection{Channel Selection}
We used two types of studies for FFT analysis. For making an RNN model with LSTM with the help of FFT processing, Emotiv Epoch+ was fitted with a total of 14 channels, which were carefully selected. The number of channels is [1,2,3,4,6,11,13,17,19,20,21,25,29,31] .The number of bands is 6. band = [4,8,12,16,25,45] . We also discovered the relation between Time domain and Frequency domain with the help of FFT in another study. 
\subsection{FFT}
Fourier Transform (FFT) is a mathematical procedure that computes the discrete Fourier transform (DFT) of a sequence. It is used to solve a variety of different types of equations or graphically depict a range of frequency activity. Fourier analysis is a signal processing technique used to convert digital signals (x) of length (N) from the timedomain to the frequency domain (X) and vice versa. FFT is a technique that is widely utilized when estimating the Power Spectral Density of an EEG signal. PSD is an abbreviation for Power spectral distribution at a specific frequency and can be computed directly on the signal using FFT or indirectly by altering the estimated autocorrelation sequence.
\subsection{RNN and LSTM}
RNNs have risen to prominence as computing power has improved, data volumes have exploded, and long short-term memory (LSTM) technology became available in the 1990s. RNNs may be incredibly precise in forecasting what will happen next because of their internal memory, which allows them to retain key input details. The reason they're so popular is because they're good at handling sequential data kinds like time series and voice. Recurrent neural networks have the advantage over other algorithms in that they can gain a deeper understanding of a sequence and its context. A short-term memory is common in RNNs. When linked with an LSTM, they have a long-term memory as well (more on that later). Due to the data sequence providing important information about what will happen next, an RNN may do jobs that other algorithms are unable to complete. \cite{b24} Long short-term memory networks (LSTMs) are a sort of recurrent neural network extension that expands memory effectively. As a result, it's well-suited to learning from big experiences separated by long periods of time. RNN extensions that increase memory capacity are known as long short-term memory (LSTM) networks. The layers of an RNN are built using LSTMs. RNNs can either assimilate new information, forget it, or give it enough importance to alter the result thanks to LSTMs, which assign “weights” to data. The layers of an RNN, which is sometimes referred to as an LSTM network, are built using the units of an LSTM. With the help of LSTMs, RNNs can remember inputs for a long time. Because LSTMs store data in a memory comparable to that of a computer, this is the case. The LSTM can read, write, and delete information from its memory. This memory can be thought of as a gated cell, with gated signifying that the cell decides whether to store or erase data (i.e., whether to open the gates) based on the value it assigns to the data. To allocate importance, weights are utilized, which the algorithm also learns. This basically means that it learns over time which data is critical and which is not. Long-Short-Term Memory Networks (LSTMs) are recurrent neural network subtypes (RNN).
\subsection{Feature Extraction}
Extracting features from EEG data can be done in a variety of methods. Periodogram and power spectral density calculations and combining band waves of various frequencies are required for feature extraction with the help of FFT.
The Welch method is \cite{b21} a modified segmentation scheme for calculating the average periodogram. Generally the Welch method of the PSD can be described by the equations below, the power spectra density, P(f) equation is defined first. Then, for each interval, the Welch Power Spectrum, $P_{\text {welch }}(f)$, is given as the mean average of the periodogram.
\begin{equation}
P(f)=\frac{1}{M U}\left|\sum_{n=0}^{M-1} x_{i}(n) w(n) e^{-j 2 \pi f}\right|^{2}
\end{equation}
\vspace{0.5cm}
\begin{equation}
P_{\text {welch }}(f)=\frac{1}{L} \sum_{i=0}^{L-1} P(f)
\end{equation}
The power spectral density (PSD) shows how a signal’s power is distributed in the frequency domain. Among the PSD estimators, Welch’s method and the multitaper approach have demonstrated the best results \cite{b22}.
The input \cite{b23} signal x [n], n = 0,1,2,…,N-1 is divided into a number of overlapping segments. Let M be the length of each segment, using n=0,1, 2,…,M-1, M.
\begin{align}
    x_i = x [i\times\frac{M}{2} + n] 
\end{align} where n=0,…,M-1,i=0,1,2,…,N-1
\newline
Each segment is given a smooth window w(n). In most cases, we employ the Hamming window at a time. The Hamming window formula for each segment is as follows:
\begin{align}
    w(n)=0.54-0.46cos[\frac{2n\pi}{M}]
\end{align}
Here, 
\begin{equation}
U=(1 / M) \sum_{n=0}^{M-1} w^{2}(n)
\end{equation}
denotes the mean power of the window w(n). 
So, \begin{equation}
M U=\sum_{n=0}^{M-1} w^{2}(n)
\end{equation}
denotes the energy of the window function w(n) with length M.
\newline
It is to be noted that, L denotes the number of data segment.

For validation, ”Accuracy” is the most popular metric. However, a model’s performance cannot be judged based only by the accuracy. So, we have used other metrics, such as - precision, recall, and f-score. The metrics were calculated using the mean of metrics for all the folds through cross validation.
\section{Results}
In our research, we tried to come up with a relation among EEG channel, time domain and frequency domain using Welch’s Periodogram with the help of band wave and FFT. The band waves identify the following emotions.
\begin{table}[ht]
\centering
\begin{tabular}{|l|l|l}
\hline Band Waves & Frequency $(\mathrm{Hz})$ & Features or emotions \\
\hline Theta & $4-8$ & Drowsiness,Mental Connection, \\
 & & Creativity \\
\hline Alpha &$8-16$ & Reflection \& Relaxation \\
\hline Beta & $16-32$ & Concentration, Problem Solving \\
\hline Gamma & $32-64$ & Learning, Perception, Multi-tasking \\
\hline
\end{tabular}
\caption{Band waves and emotions}
\end{table}
\newline
The following figure shows the time domain of the EEG signals. From the figure, we can see that there has been lots of electrical activities going on the EEG channels. 
\begin{figure}[htbp]
\centerline{\includegraphics[width=9cm]{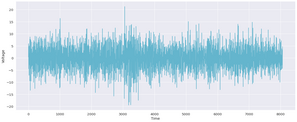}}
\caption{Time Domain of the EEG Signals}
\label{fig}
\end{figure}
And from the time domain, we can get the graphs of the frequency domain along with Power Spectral Density across the channels with the help of Fourier Transformation. In our study, we used Fast Fourier Transformation, the sine wave was taken from 4 Hz to 45 Hz. So, by comparing the sine wave with the time domain, we can get the PSD at the frequency domain. From the time-frequency domain, we can see the electrical activity in brain in multiple time intervals which shows a relation between different frequencies, brain activity and voltage.
\vspace{0.4cm}
\begin{figure}[htbp]
\centerline{\includegraphics{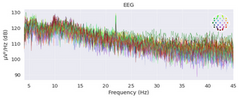}}
\caption{Power Spectral Density Across Channels}
\label{fig}
\end{figure}
\newline
For the first FFT analysis, the research calculates mean, std, min, first quartile, median, third quartile and max values of 1240 trials of the six regions based on sensors and four band power values. For this research, we used SVM and K-NN classifiers. SVM classifier used “linear” kernel in this research. The research also calculates the accuracy of Valence and Arousal.
\begin{table}[ht]
\centering
\begin{tabular}{|c|c|c|}
\hline Accuracy in \% & SVM & K-NN \\
\hline Arousal & $58.52 \%$ & $62.32$ \\
\hline Valence & $56.79 \%$ & $56.92 \%$ \\
\hline
\end{tabular}
\caption{Accuracy of Arousal and Valence using SVM and KNN Classifiers}
\end{table}
\newline
We attempted to experience the variations in electrical activities in the brain over time in the first study. To extract EEG signals, the 32 sensor sites were separated into globally recognizable zones. The position of the electrode are frontal, central, temporal, parietal, and occipital placements, respectively.
The topographical maps are used to visualize spatial distribution of activity. This useful visualization method allows us to examine how data changes over one time point to another. The subject in this study was watching a video while we analyzed the changes in electrical activity from 0.153 to 0.273 seconds. We can see the changes of electrical activity in voltage based  various frequencies as band waves are determined by the range of frequency and different band waves indicate different ranges of emotion. From this, it can be said that the subject can feel different emotions in a particular time point. 
\begin{figure}[htbp]
\centerline{\includegraphics{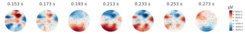}}
\caption{Topographical Map for Theta band wave}
\label{fig}
\end{figure}
\begin{figure}[htbp]
\centerline{\includegraphics{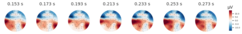}}
\caption{Topographical Map for Alpha band wave}
\label{fig}
\end{figure}
\begin{figure}[htbp]
\centerline{\includegraphics{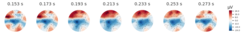}}
\caption{Topographical Map for Beta band wave}
\label{fig}
\end{figure}
\begin{figure}[htbp]
\centerline{\includegraphics{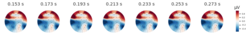}}
\caption{Topographical Map for Gamma band wave}
\label{fig}
\end{figure}
\newline
For the second research, during the FFT processing, we employed meta data for the purpose of doing a meta vector analysis. Raw data was split over a time span of 2 seconds, with each slice having a 0.125-second interval between it. A two-second FFT of channel was carried out in different frequencies in a sequence. Emotiv Epoch+ was fitted with a total of 14 channels, which were carefully selected. The number of channels is [1,2,3,4,6,11,13,17,19,20,21,25,29,31] .The number of bands is 6. band = [4,8,12,16,25,45] . A band power of 2 seconds on average is used. The window size was 256 with a step size of 16, with each update occurring once every 0.125 seconds. The sampling rate was set to 128 hertz. The FFT was then performed on all of the subjects using these settings in order to obtain the required output.
Neural networks and other forms of artificial intelligence require a starting collection of data, referred to as a training dataset, that serves as a foundation for subsequent application and use. This dataset serves as the foundation for the program's developing information library. Before the model can interpret and learn from the training data, it must be appropriately labeled.
The lowest value of the data is 200 and the greatest value is above 2000, which means that trying to plot it will result in a lot of irrelevant plots, which will make conducting the analysis tough. The objective of machine learning is to create a plot and then optimize it further in order to obtain a pattern. And if there are significant differences between the plotted points, it will be unable to optimize the data. As a result, in order to fix this issue, the values have been reduced to their bare minimum, commonly known as scaling. The values of the data will not be lost as a result of scaling; instead, the data will be optimized to the point where there is little difference between the plotted points. 
In order to achieve this, StandardScaler must transform your data into a distribution with a mean of zero and a standard deviation of one. When dealing with multivariate data, this is done feature-by-feature to ensure that the data is accurate (in other words independently for each column of the data). Because of the way the data is distributed, each value in the dataset will be deducted from the mean and then divided by the standard deviation of the dataset.
After that, we divided the data set into two parts: a training data set and a testing data set. Training will be carried out on 75\% of the data, and testing will be carried out on 25\% of the data. A total of 456768 data were used in the training process. A total of 152256 data were used in the testing.
RNN has been kept sequential. The first layer LSTM of sequential model takes input of 512. The second layer takes input of 256. The third and fourth layer takes an input of 128 and 64. And, the final layer LSTM of sequential model takes input of 10. Since we are conducting classification where we will need 0 or 1 that is why sigmoid has been used. The activation functions used are relu and for the last part sigmoid. The rectified linear activation function, abbreviated ReLU, is a piecewise linear function that, if the input is positive, outputs the value directly; otherwise, it outputs zero.
Batch normalization was used. Batch normalization is a method for training extremely deep neural networks in which the inputs to a layer are standardized for each mini-batch. This results in a stabilization of the learning process and a significant drop in the total of training epochs required for training deep networks. Through randomly dropping out nodes while training, a single model can be utilized to simulate having a huge variety of distinct network designs.[2] This is referred to as dropout, and it is an extremely computationally efficient and amazingly successful regularization technique for reducing overfitting and improving generalization error in all types of deep neural networks. In our situation, dropout rates began at 30\%, increased to 50\%, then 30\%, 30\%, 30\%, and eventually 20\%.
We worked with three-dimensional datasets; however, when we converted to a dense layer, we obtained a one-dimensional representation in order to make a prediction. RMSprop was used as the optimizer with a learning rate of 0.001, a rho value of 0.9, and an epsilon value of 1e-08. RMSprop calculates the gradient by dividing it by the root of the moving (discounted) average of the square of the gradients. This application of RMSprop makes use of conventional momentum rather than Nesterov momentum. Additionally, the centered version calculates the variance by calculating a moving average of the gradients. As we can see, accuracy increases very gradually in this case, and learning rate plays a major part. If we increased the learning rate, accuracy would also increase rapidly, and when optimization is reached, the process would reverse, with accuracy decreasing at a faster rate. That is why the rate of learning has been reduced. When one zero is removed, the accuracy decreases significantly.
As our loss function, we utilized the Mean Squared Error. The Mean Squared Error (MSE) loss function is the most basic and extensively used loss function, and it is typically taught in introductory Machine Learning programs. To calculate the MSE, take the difference between your model's predictions and the ground truth, square it, and then average it across the whole dataset. The MSE can never be negative since we are constantly squaring the errors. To compute loss, we utilized mean squared error. Because of the squaring portion of the function, the MSE is excellent for guaranteeing that the trained model does not contain any outlier predictions with significant mistakes. Because of this, the MSE places greater emphasis on outlier predictions with large errors.
We tried our best to reduce the percentage of value loss and increase the accuracy rate. We saved the model and kept track by every 50 epochs. In the first picture, we can see that for the first 50 epochs the training loss 0.1588 and validation loss reduced to 0.06851 and 0.06005. And the training accuracy rate increased from 9.61 percent to 45.784 percent and validation accuracy increased to 53.420 pecent. For the second 50 epochs, the training loss reduced to 0.06283 and the validation loss reduced to .05223 where the training accuracy increased to 51.661 percent and validation accuracy increased to 60.339 percent. For the third 50 epochs, the training loss reduced to 0.05992 and the validation loss reduced to .04787 where the training accuracy increased to 54.492 percent and validation accuracy increased to 64.413 percent. After 200 epochs the ratio started to change at a very slow rate.We ran 1000 epochs and got the training accuracy rate of 69.21\% and the validation accuracy rate was 78.28\%.
\section{Conclusion}
To summarize, in this research, we describe the EEG-based emotion recognition challenge, as well as existing and proposed solutions to this problem. Emotion detection by the use of EEG waves is a relatively new and exciting area of study and analysis. To identify and evaluate on numerous emotional states using EEG signals acquired from the DEAP Dataset, SVM (Support Vector Machine), KNN (K-Nearest Neighbor). According to the findings, the suggested method is a very promising option for emotion recognition, owing to its remarkable ability to learn features from raw data in a short period of time. When compared to typical feature extraction approaches, it produces higher average accuracy over a larger number of people.

\end{document}